\documentclass[oneside,a4paper,draft,reqno]{amsart}
\usepackage{amsmath}
\usepackage{xspace}
\usepackage[mathscr]{eucal}

\newcommand{\mat}[3]{\ensuremath{
																								\left \langle  \vphantom{#2 #3}   #1   
                        \right|    					\, #2\,   
                        \left|    \vphantom{#2 #1} #3   
                        \right \rangle
                        								}
                     }
\newcommand{\bmat}[3]{\ensuremath{
																								\bigl \langle     #1   \bigr|    \, #2\,   \bigl|     #3   \bigr \rangle
                        									}
                     }
 
\newcommand{\bsubmatB}[5]{\ensuremath{     {\vphantom{\bket{#1}}}_{#4} \mspace{- #5 mu} 
																																														 \bmat{#1}{#2}{#3}
																																								}
                           }
\newcommand{\ket}[1]{\ensuremath{		\left| #1 \right> 
																																			  }
																									}
\newcommand{\bket}[1]{\ensuremath{		\bigl| #1 \bigr> 
																																			  }
																									}
 
\newcommand{\bbra}[1]{\ensuremath{		\bigl< #1 \bigr| 
																															  }
																					}
\newcommand{\coh}[3]{\ensuremath{		\left( #1, #2 \right)_{#3} 
																																			  }
																									}

\newcommand{\overlap}[2]{\ensuremath{ 
																								\left \langle    #1 \vphantom{#2 } \,
                        \right| \left.   #2 \vphantom{#1}
                        \right \rangle
                        									}
                     }
\newcommand{\boverlap}[2]{\ensuremath{ 
																								\bigl \langle #1 \, \bigr| \bigl. #2 \bigr \rangle
                        									}
                     }
      
\newcommand{\bsuboverlapB}[4]{\ensuremath{     {\vphantom{\bket{#1}}}_{#3} \mspace{- #4 mu} 
																																														 \boverlap{#1}{#2} 
																																								}
                           }
\newcommand{\comm}[2]{\ensuremath{  \left[ #1, #2 \right] }}
\newcommand{\bcomm}[2]{\ensuremath{							\bigl[ #1 , #2 \bigr]							}}
\newcommand{\pt}{   \ensuremath{     \phi_{\mathrm{ap}}     }}
\newcommand{\xOp}{ \ensuremath{  \hat{x}  }}
\newcommand{\pOp}{ \ensuremath{  \hat{p}  }}
\newcommand{\xiOp}{ \ensuremath{  \hat{x}_{\mathrm{i}}  }\xspace}
\newcommand{\piOp}{ \ensuremath{  \hat{p}_{\mathrm{i}}  }\xspace}
\newcommand{\xfOp}{ \ensuremath{  \hat{x}_{\mathrm{f}}  }\xspace}
\newcommand{\pfOp}{ \ensuremath{  \hat{p}_{\mathrm{f}}  }\xspace}
\newcommand{\MxOp}{  \ensuremath{   \hat{\mu}_{\mathrm{X}}    } }
\newcommand{\MpOp}{  \ensuremath{   \hat{\mu}_{\mathrm{P}}    } }

\newcommand{\MxfOp}{  \ensuremath{   \hat{\mu}_{\mathrm{Xf}}    } }
\newcommand{\MpfOp}{  \ensuremath{   \hat{\mu}_{\mathrm{Pf}}    } }
\newcommand{\Mx}{   \ensuremath{   \mu_{\mathrm{X}}  } }
\newcommand{\Mp}{   \ensuremath{   \mu_{\mathrm{P}}  } }
\newcommand{\PxOp}{  \ensuremath{   \hat{\pi}_{\mathrm{X}}    } }
\newcommand{\PpOp}{  \ensuremath{   \hat{\pi}_{\mathrm{P}}    } }

\newcommand{\PxfOp}{  \ensuremath{   \hat{\pi}_{\mathrm{Xf}}    } }

\newcommand{\Px}{   \ensuremath{   \pi_{\mathrm{X}}  } }
\newcommand{\Pp}{   \ensuremath{   \pi_{\mathrm{P}}  } }
\newcommand{\Exi}{\ensuremath{       \hat{\epsilon}_{\mathrm{Xi}}       }\xspace}
\newcommand{\Epi}{\ensuremath{       \hat{\epsilon}_{\mathrm{Pi}}       }\xspace}
\newcommand{\Ei}{\ensuremath{        \epsilon_{\mathrm{i}}       }\xspace}
\newcommand{\Exf}{\ensuremath{       \hat{\epsilon}_{\mathrm{Xf}}       }\xspace}
\newcommand{\Epf}{\ensuremath{       \hat{\epsilon}_{\mathrm{Pf}}       }\xspace}
\newcommand{\Ex}{\ensuremath{       \hat{\epsilon}_{\mathrm{X}}       }\xspace}
\newcommand{\Ep}{\ensuremath{       \hat{\epsilon}_{\mathrm{P}}       }\xspace}
\newcommand{\Ef}{\ensuremath{        \epsilon_{\mathrm{f}}       }\xspace}
\newcommand{\Mxf}{   \ensuremath{   \mu_{\mathrm{Xf}}  } }
\newcommand{\Mpf}{   \ensuremath{   \mu_{\mathrm{Pf}}  } }
\newcommand{\Exiv}{\ensuremath{       \epsilon}_{\mathrm{Xi}       }\xspace}
\newcommand{\Epiv}{\ensuremath{       \epsilon}_{\mathrm{Pi}       }\xspace}
\newcommand{\Exfv}{\ensuremath{       \epsilon}_{\mathrm{Xf}       }\xspace}
\newcommand{\Epfv}{\ensuremath{       \epsilon}_{\mathrm{Pf}       }\xspace}
\newcommand{\Dx}{\ensuremath{       \hat{\delta}_{\mathrm{X}}       }\xspace}
\newcommand{\Dp}{\ensuremath{       \hat{\delta}_{\mathrm{P}}       }\xspace}
\newcommand{\RErr}{\ensuremath{     \Delta_{\mathrm{ei}}  }\xspace}
\newcommand{\PErr}{\ensuremath{     \Delta_{\mathrm{ef}}  }\xspace}
\newcommand{\Dist}{\ensuremath{     \Delta_{\mathrm{d}}  }\xspace}
\begin{document}
\begin{titlepage}
\begin{center}
\bfseries
MAXIMAL ACCURACY AND MINIMAL DISTURBANCE IN THE ARTHURS-KELLY SIMULTANEOUS MEASUREMENT PROCESS
\end{center}
\vspace{1 cm}
\begin{center}
D M APPLEBY
\end{center}
\begin{center}
Department of Physics, Queen Mary and
		Westfield College,  Mile End Rd, London E1 4NS, UK
 \end{center}
\vspace{0.5 cm}
\begin{center}
  (E-mail:  D.M.Appleby@qmw.ac.uk)
\end{center}
\vspace{1.25 cm}
\begin{center}
\textbf{Abstract}\\
\vspace{0.35 cm}
\parbox{10.5 cm }{   The accuracy of the Arthurs-Kelly model of a simultaneous measurement of position and  
                     momentum is analysed using  concepts developed by Braginsky and Khalili in the context
                     of measurements of a single quantum observable.  A distinction is made between the errors
                     of retrodiction and prediction.  It is shown that the distribution of measured values 
                     coincides with the initial state Husimi
                     function when the retrodictive accuracy is maximised
                     (this being the result first obtained by Busch, 
                     using somewhat different methods).  It is further shown that the distribution
                     of measured values is
                     related to the 
                     final state anti-Husimi function (the $P$ representation of quantum optics) when
                     the predictive accuracy is maximised.  The disturbance of the system by the 
                     measurement is also discussed.  A class of
                     minimally disturbing measurements is characterised.  
                     It is shown that the distribution of measured values then coincides
                     with one of the smoothed Wigner functions described by Cartwright.
                 }\\     
\vspace{1.0 cm}
\parbox{10.5 cm }{
}
\end{center}
\vspace{1 cm}
\begin{center}
Report no.  QMW-PH-98-12
\end{center}
\end{titlepage}
\section{Introduction}
\label{sec:  intro}
It has been known since the publication of Heisenberg's original
paper\cite{Heis1} that quantum mechanics does not allow both the
position and the momentum of a system to be measured with
arbitrary accuracy.   However, it does not follow from this that
one cannot measure both quantities with a less than
perfect degree of accuracy.  Indeed, it would seem that it is
essential that quantum mechanics does permit
such measurements, if it is to be possible to
derive classical mechanics from quantum
mechanics as an approximate theory, valid in
some appropriate limit.  For this and other
reasons simultaneous measurements have been the
subject of much theoretical interest over the
years.

In recent years interest in them has been greatly increased,
due to technical advances in the field of quantum optics.  As a
result of these advances simultaneous, imperfect measurements of
the position and momentum of a quantum mechanical system
are no longer confined to the idealised
world of 
\emph{gedanken} experiments.  They can actually be realised in the
laboratory~\cite{ArthursKelly,Braunstein,Stenholm,Leon1,LeonUncert,Torma,Power,Bus4}.  
For a recent
review of these methods, and of the closely related subject
of the tomographic reconstruction of the quantum state, the
reader may consult Leonhardt and Paul~\cite{Leon2},
or Leonhardt~\cite{Leon3}.

In the following we shall be concerned with the
problem of describing the accuracy of such
measurements, and  the disturbance they produce
in the system whose  position and momentum are
being measured.  In a previous paper~\cite{Appleby} we developed some general
methods for analysing this problem.  Our purpose now is to 
illustrate the use of these methods, by applying them to
the example of the
Arthurs-Kelly process~\cite{ArthursKelly,Braunstein,Stenholm,Bus4,Leon3}.

The accuracy of, and disturbance produced by a simultaneous measurement
process has been the subject of many previous investigations.
One important relationship is the inequality first 
derived by Arthurs and Kelly~\cite{ArthursKelly,LeonUncert,Leon3,WodArtKel,Raymer}
\begin{equation}
  \Delta \Mxf  \,  \Delta  \Mpf \ge \hbar
\label{eq:  ArtKelRel}
\end{equation}
where $\Delta \Mxf$ and $\Delta \Mpf$
denote the uncertainties in the final pointer positions, after the
measurement has been completed.  This inequality is to be compared
with the ordinary uncertainty principle
\begin{equation}
  \Delta x \, \Delta p \ge \frac{\hbar}{2}
\label{eq:  UncertRel}
\end{equation}
The doubling of the lower bound in Eq.~(\ref{eq:  ArtKelRel}) as 
compared with Eq.~(\ref{eq:  UncertRel}) is due to  the additional
noise introduced by the measurement process.

We  discussed the Arthurs-Kelly inequality in ref.~\cite{Appleby}.  We
will here content ourselves with remarking, that although
the product $\Delta \Mxf \, \Delta  \Mpf$
provides  a numerical indication of the overall accuracy of the 
measurement, the quantities $\Delta \Mxf$ and $\Delta \Mpf$ are not
themselves directly interpretable as experimental errors. 
One would like to find quantities which can be individually 
and immediately identified with
the errors in the measured values of position and momentum.

The problem of defining quantities which do provide a direct 
characterisation of the errors in
a simultaneous measurement of position and momentum has been 
systematically addressed 
within the framework of the approach based on the
concepts of a positive operator valued measure (or POVM) and an unsharp 
observable~\cite{Prug1,Davies,Prug2,Prug2b,
Prug3,Hol,Prug4,Bus1,Bus2,Bus3,
Mart1,Mart2,Mart3,Ban}.  For a recent review, and additional references, the reader may
consult the book by Busch \emph{et al}~\cite{Bus4}.  We give a brief discussion
of the way in which the measurement errors are defined in this approach
in the Appendix.

Our own interest in this problem stems from the work of Uffink~\cite{Uffink},
who has criticised the POVM approach, and more generally the very
concept of a joint measurement of non-commuting observables, in terms which may fairly be 
described as unequivocal.  He states
\begin{quote}
  \dots the formalism of quantum theory, as it
is presented by von Neumann, simply has no room
for a description of a joint measurement of
position and momentum at all. [p.200]
\end{quote}
He further asserts
\begin{quote}
 the claim that within this formalism
[\emph{i.e.}\ the formalism based on 
on the concept of a POVM] a joint
unsharp measurement of position and momentum 
or a pair of spin components is possible is
false.
\end{quote}
He goes on to assert, that  claims that the POVM approach does 
provide a theory of joint measurements, 
\begin{quote}
  rest on the adoption of inappropriate definitions, 
  \emph{i.e.}, definitions that trivialize the problem.
\end{quote}
In connection with these statements, we should say at the outset, 
that we do not feel that Uffink's arguments are sufficiently strong
to support his conclusion.
On a purely
intuitive level, it seems evident that the
processes discussed in 
refs.~\cite{ArthursKelly,Braunstein,Stenholm,
Leon1,LeonUncert,Torma,Power,Bus4,Leon2,Leon3} 
must be describable as joint measurements
in some sense of the word.  It seems no less evident, that it cannot
really be correct to dismiss the claims of the POVM approach in the
way that Uffink does dismiss them, wholesale, 
as  ``false'' and/or ``trivial''.  
In short, it appears to us that Uffink overstates his case.
However, that is not to say that his criticisms are without foundation.
On the contrary, we consider
that Uffink has raised some very important and interesting questions of principle which 
it would be worthwhile trying to clarify. This was the original motive for the work reported
in ref.~\cite{Appleby}.

If one examines the statements just quoted it will be perceived that Uffink
actually makes two different claims, which may be enumerated
as follows:
\begin{enumerate}
\item  The conventional, generally accepted formulation of quantum mechanics, as it is
       presented in the book by Von Neumann~\cite{VonNeu}, has no room for the concept
       of a simultaneous measurement of position and momentum.  Or, to put it another way,
       the conventional theory of quantum mechanics does not permit both
       the position and momentum of a system to be jointly determined, not even 
       with a less than perfect degree of accuracy.
\item  The attempt to solve the joint measurement problem within the framework of
       the approach based on POVM's and unsharp observables fails, because
       it depends on definitions which are ``inappropriate'' because
       they ``trivialize'' the problem.
\end{enumerate}
In ref.~\cite{Appleby} we addressed the first of these objections
(we intend to address the second objection in a future publication).
We took it, that a measurement consists of a process in which a correlation
is established between one set of observables (the quantities being measured)
and another set of observables (the meter readings).  In the case of interest
to us, the observables being measured are the position $\xOp$ and momentum
$\pOp$ of a quantum mechanical system having one degree of freedom.  The observables
representing the measurement outcome are two commuting pointer observables
$\MxOp$ and $\MpOp$.  A process which brings about, simultaneously,
both a correlation between $\xOp$ and $\MxOp$, and a correlation between
$\pOp$ and $\MpOp$, is \emph{ipso facto}  describable as a 
simultaneous measurement of $\xOp$ and $\pOp$.  The problem then reduces
to the problem of giving a precise characterisation of the degree of 
this correlation.  In other words, it reduces to the problem of giving
a precise characterisation of the measurement accuracy.
In ref.~\cite{Appleby} we showed how this problem 
can be solved by taking some concepts developed by Braginsky 
and Khalili~\cite{Braginsky} in 
the context of single measurements of position only, and applying them 
to the case of simultaneous measurements of position and momentum
together.  At no stage did we go beyond the conventional theory 
of quantum mechanics, as it is presented by Von Neumann.  We were therefore 
able to conclude,
that contrary to what Uffink asserts, the conventional theory
does permit the simultaneous, albeit less
than perfectly accurate, measurement of both the position and the
momentum of a quantum mechanical system.

Our motive for undertaking the research reported in ref.~\cite{Appleby}
was the desire to clarify a fundamental question of principle.
However, in the course of this research it became clear to us that
the methods we had developed had an interest which went somewhat
beyond the purpose for which we had originally intended them.
It appears to us that they provide one with a way of analysing 
the physics of simultaneous
measurement processes which is useful, both as a mathematical
technique, and as a source of physical insight. 

We discussed the fundamental concepts on 
which our approach in based in ref.~\cite{Appleby}
(also see Section~\ref{sec:  ErrDis} below).
Our approach rests on making a careful distinction between
the retrodictive and predictive (or determinative and preparative) aspects
of a joint measurement process.  Corresponding to this distinction we 
introduce two different kinds of error:  a pair of retrodictive errors
and a pair of predictive errors.  We also define a pair of disturbances, 
providing a quantitative indication of the disturbance of the
system by the measurement process.  
We then derive a total of six inequalities relating these quantities:
\emph{viz} a retrodictive error relationship, a predictive error relationship,
and four error-disturbance relationships.
As we stated above, our approach
represents a development of the analysis
given by Braginsky and Khalili~\cite{Braginsky} 
for the case of single measurements of position only.

In the case of joint measurement processes
Ali and Prugove\v{c}ki~\cite{Prug2b}
give a definition which is essentially equivalent to our definition
of the errors of retrodiction for the class of processes
to which their definition is applicable.  In particular, our definition is
essentially equivalent to theirs in the case of the Arthurs-Kelly process,
which forms the subject of this paper.
However, it can be shown (see the Appendix) that our definition 
is the more general of the two, in that it applies to a class of measurement
processes which is strictly larger than the class considered
by Ali and Prugove\v{c}ki.  Also, our definition provides 
additional physical insight even in those cases 
where it is equivalent to---gives the same 
numerical values as---the definition of Prugove\v{c}ki
and Ali.  We used this fact in ref.~\cite{Appleby}, when trying to
answer Uffink's objection, that the conventional theory of
quantum mechanics, as formulated by von Neumann, has no room
for the concept of a simultaneous measurement of non-commuting
quantities.

However, the most important  feature of our approach is the way in which
we analyse the distinction between the retrodictive and predictive
aspects of a joint measurement process, and the way in which we
relate these two aspects to the disturbance produced by the
measurement.  The distinction between retrodiction and prediction
has been discussed by several other authors,
as has the reciprocity between the accuracy of a joint
measurement, and the disturbance 
produced~\cite{Bus4,Prug1,Bus1,Bus2,Bus3,Mart2,Margenau,Hilgevoord}.
However, it appears to us that our method of analysis is a significant
source of additional insight.  It also has certain technical
advantages, in that it provides a numerical characterisation of the
various contrasting features of a joint measurement process which is 
particularly concise, and convenient.

In the following sections we will illustrate these points,
with an analysis of 
the Arthurs-Kelly process.  This process has been
the 
subject of several previous 
investigations~\cite{ArthursKelly,Braunstein,Stenholm,Power,Bus4,Leon3,Bus2}. 
We derive a number of new results. 
However, in some cases we only give an alternative derivation of a result which is already
known. We thought that this was worth doing since our primary purpose is to illustrate
the use of some general methods, which have an application to many other
measurement processes.

We begin, in Section~\ref{sec:  ErrDis}
by giving a brief summary of the relevant results from
ref.~\cite{Appleby}.  In 
section~\ref{sec:  DistMeasVal}, we show how
the initial apparatus state may
conveniently be expanded in terms of
eigenstates of the retrodictive and
predictive error operators.  We then
use this fact to show how the distribution of
measured values  depends on the distribution of
retrodictive errors; and how the final state of
the system depends on the distribution of
predictive errors.  In particular, by
appropriately choosing the distribution of
retrodictive errors, it is possible to ensure
that the distribution of measured values is
given by any member of the class of ``operational
distributions'' which are obtained by taking the convolution of 
two Wigner functions, as discussed
by Davies~\cite{Davies}, W\'{o}dkiewicz~\cite{Operational} 
and others~\cite{Bus4,Ban,Lalovic,Halliwell}.  With
the appropriate choice for the distribution of
predictive errors it is possible to prepare the
system in any desired pure state.  

In
section~\ref{sec:  RetOpt} we consider 
retrodictively optimal
processes:  \emph{i.e.} processes which minimise
the product of retrodictive errors. We show
that in every such case the distribution of
results is given by the initial system state
Husimi function~\cite{Husimi,Reviews} (or $Q$
representation).  This is the result first obtained
by Ali and Prugove\v{c}ki~\cite{Prug2b} and,
in the context of the Arthurs-Kelly process, by 
Busch~\cite{Bus2}.

In section~\ref{sec:  PreOpt} we consider
predictively optimal processes, which minimise
the product of predictive errors.  We show that
the distribution of measured values is then
related to the final system 
state anti-Husimi
function~\cite{Reviews,Glauber}  (or $P$
representation).

The distributions of retrodictive and
predictive errors are  independent of one
another:  it is possible for a measurement to be
optimal retrodictively, whilst being very poor
predictively, or \emph{vice versa}.  In 
section~\ref{sec:  complete} we consider
completely optimal processes:  \emph{i.e.}
processes which are both retrodictively and
predictively optimal, and which also minimise
the degree of disturbance.

Finally, in section~\ref{sec:  MinDis}, we
consider the disturbance of the system by the
measurement.  It is possible to reduce the
level of disturbance below that produced by an
optimal measurement, provided one is willing to
accept a reduced degree of accuracy.  In
section~\ref{sec:  MinDis} we consider
measurements which give the maximum degree of
accuracy consistent with a given level of
disturbance.

Before concluding these introductory remarks we should, perhaps,
observe that the contrast we drew above,
between the ``POVM approach,'' and the approach 
adopted in the present paper, is potentially misleading.  
It is, in fact, almost impossible to
talk about simultaneous measurement processes without tacitly introducing
the concept of a POVM.  When we distinguish
our own approach from the POVM approach we do not mean to suggest 
that POVM's are not implicitly present in our analysis.
We only mean that the concept does not explicitly play the same central role
that it does in, for example, the book by Busch \emph{et al}~\cite{Bus4}.

In the following pages we make no direct use of results obtained within
the context of the POVM approach.  That is because the work reported here
is a continuation of the work reported in ref.~\cite{Appleby}.
The original motive for this work
was our desire to answer the first of Uffink's objections (as 
itemised above).  In order to do so we considered it to be necessary
to go back to first principles, and to think things through from
the foundation up.  We were thereby led to a different approach
to the theory of simultaneous measurements of position and momentum, which
it seems most natural to present in the manner in which we discovered 
it, independently of the POVM approach.

It would obviously be desirable to relate the approach taken in this paper
to the POVM approach.  However, that is a task which we prefer to leave
to a future publication, when we also hope to address the second of 
Uffink's objections (as itemised above).
\section{Definition of the Errors and
Disturbances}
\label{sec:  ErrDis}
In the process described by Arthurs and Kelly~\cite{ArthursKelly} a system, 
with position $\xOp$ and momentum $\pOp$, interacts with an apparatus
characterised by two pointer observables $\MxOp$ (measuring the value of $\hat{x}$)
and $\MpOp$ (measuring the value of $\hat{p}$).  
Let $\PxOp$ and $\PpOp$ be the momenta conjugate to $\MxOp$ and $\MpOp$ respectively.
Then
\begin{equation*}
  \comm{\xOp}{\pOp} = \comm{\MxOp}{\PxOp} = \comm{\MpOp}{\PpOp} = i \hbar
\end{equation*}
all other commutators between 
the operators $\xOp$, $\pOp$, $\MxOp$, $\PpOp$, $\MpOp$, $\PpOp$ being zero.

The measurement interaction is described by the unitary evolution operator
\begin{equation}
 \hat{U} = \exp \bigl[- \tfrac{i}{\hbar} \left( \PpOp \pOp + \PxOp \xOp \right)\bigr]
\label{eq:  HatUDef}
\end{equation}
We assume that  system+apparatus are initially 
in the product state $\ket{\psi \otimes \pt}$,
$\ket{\psi}$ being the initial state of the system, and $\ket{\pt}$ the initial state of the apparatus.

In order to define the errors and disturbances we switch to the Heisenberg picture.
Let $\hat{\mathscr{O}}$ be any of the Schr\"{o}dinger picture 
operators $\xOp$, $\pOp$, $\MxOp$, $\PpOp$, $\MpOp$, $\PpOp$.
Let $\hat{\mathscr{O}}_{\mathrm{i}}=\hat{\mathscr{O}}$ be the
value of the corresponding Heisenberg picture operator immediately before the interaction.
Let $\hat{\mathscr{O}}_{\mathrm{f}} = \hat{U}^{\dagger} \hat{\mathscr{O}} \hat{U}$
be its value immediately afterwards.  Then
\begin{equation}
\begin{aligned}
  \xfOp  & = \xOp + \PpOp                         & \hspace{0.5 in} \pfOp  & = \pOp - \PxOp  \\
  \MxfOp & = \MxOp + \xOp + \tfrac{1}{2} \PpOp    & \hspace{0.5 in} \PxfOp & = \PxOp         \\
  \MpfOp & = \MpOp + \pOp - \tfrac{1}{2} \PxOp    & \hspace{0.5 in} \PxfOp & = \PxOp        
\end{aligned}
\label{eq:  FinHeisOps}
\end{equation}
Define the operators $\Exi$, $\Epi$, $\Exf$, $\Epf$, $\Dx$, $\Dp$ by
\begin{equation}
\begin{aligned}
  \Exi & = \MxfOp - \xiOp  & \hspace{0.5 in} \Epi & = \MpfOp - \piOp \\
  \Exf & = \MxfOp - \xfOp  & \hspace{0.5 in} \Epi & = \MpfOp - \pfOp \\
  \Dx  & = \xfOp - \xiOp   & \hspace{0.5 in} \Dp  & = \pfOp - \piOp
\end{aligned}
\label{eq:  ErrDisOps}
\end{equation}
Following Braginsky and Khalili~\cite{Braginsky} we then define the
rms errors of retrodiction
\begin{equation}
  \RErr x = \Bigl(\bmat{\psi \otimes \pt}{\Exi^2}{\psi \otimes \pt} \Bigr)^{\frac{1}{2}}
  \hspace{0.5 in}
  \RErr p = \Bigl(\bmat{\psi \otimes \pt}{\Epi^2}{\psi \otimes \pt}\Bigr)^{\frac{1}{2}} 
\label{eq:  RetErrDef}
\end{equation}
and the rms disturbances
\begin{equation*}
  \Dist x =  \Bigl(\bmat{\psi \otimes \pt}{\Dx^2}{\psi \otimes \pt}\Bigr)^{\frac{1}{2}} 
  \hspace{0.5 in}
  \Dist p = \Bigl(\bmat{\psi \otimes \pt}{\Dp^2}{\psi \otimes \pt}\Bigr)^{\frac{1}{2}} 
\end{equation*}
$\RErr x$, $\RErr p$ correspond to the quantities 
$\Delta x_{\mathrm{measure}}$, $\Delta p_{\mathrm{measure}}$ defined by
Braginsky and Khalili in the context of single measurements of $\xOp$ or $\pOp$ only.
They provide a numerical indication of the accuracy with which the result of the measurement reflects
the initial state of the system.
$\Dist x$, $\Dist p$ correspond to  Braginsky and Khalili's 
$\Delta x_{\mathrm{perturb}}$, $\Delta p_{\mathrm{perturb}}$.  They provide a numerical
indication of the degree to which the measurement perturbs the state of the system.

We also define the rms errors of prediction
\begin{equation*}
  \PErr x = \Bigl(\bmat{\psi \otimes \pt}{\Exf^2}{\psi \otimes \pt}\Bigr)^{\frac{1}{2}} 
  \hspace{0.5 in}
  \PErr p = \Bigl(\bmat{\psi \otimes \pt}{\Epf^2}{\psi \otimes \pt}\Bigr)^{\frac{1}{2}} 
\end{equation*}
These quantities provide a numerical indication of the accuracy with which the result of the measurement
reflects the final state of the system.  Braginsky and Khalili do not consider this second
kind of error.

We have given a detailed discussion of the interpretation of the quantities 
$\RErr x$, $\RErr p$, $\PErr x$, $\PErr p$, $\Dist x$, $\Dist p$ in ref.~\cite{Appleby}.

In view of Eqs.~(\ref{eq:  FinHeisOps}) and~(\ref{eq:  ErrDisOps}) we have
\begin{equation}
\begin{aligned}
  \Exi & = \MxOp + \tfrac{1}{2} \PpOp & \hspace{0.5 in} \Epi & = \phantom{-} \MpOp - \tfrac{1}{2} \PxOp \\
  \Exf & = \MxOp - \tfrac{1}{2} \PpOp & \hspace{0.5 in} \Epf & = \phantom{-} \MpOp + \tfrac{1}{2} \PxOp \\
  \Dx  & = \PpOp                      & \hspace{0.5 in} \Dp  & = -\PxOp
\end{aligned}
\label{eq:  ErrDisExp}
\end{equation}
Consequently
\begin{equation*}
\begin{aligned}
   \bcomm{\Exi}{\Epi} & = - i \hbar           & \hspace{0.5 in} \bcomm{\Exf}{\Epf} & = \phantom{-} i \hbar \\
   \bcomm{\Exi}{\Dp}  & = - i \hbar           & \hspace{0.5 in} \bcomm{\Exf}{\Dp}  & = - i \hbar \\
   \bcomm{\Dx}{\Epi}  & = - i \hbar           & \hspace{0.5 in} \bcomm{\Dx}{\Epf}  & = - i \hbar
\end{aligned}
\end{equation*}
all other commutators between the operators
$\Exi$, $\Epi$, $\Exf$, $\Epf$, $\Dx$, $\Dp$ being zero.  We deduce the retrodictive 
and predictive error relationships
\begin{align}
  \RErr x \,\RErr p & \ge \frac{\hbar}{2}
\label{eq:  RetErrRelB}
\\
  \PErr x \, \PErr p & \ge \frac{\hbar}{2}
\label{eq:  PreErrRelB}
\end{align}
and the four error-disturbance relationships
\begin{equation}
\begin{aligned}
  \RErr x \, \Dist p & \ge \frac{\hbar}{2} & \hspace{0.5 in} \PErr x \, \Dist p & \ge \frac{\hbar}{2} \\
  \RErr p \, \Dist x & \ge \frac{\hbar}{2} & \hspace{0.5 in} \PErr p \, \Dist x & \ge \frac{\hbar}{2}
\end{aligned}
\label{eq:  ErrDisRelB}
\end{equation}
Eqs.~(\ref{eq:  RetErrRelB}) and~(\ref{eq:  PreErrRelB}) jointly comprise a precise, quantitative
statement of the well-known principle, that the product of the errors in a simultaneous measurement
of position and momentum must be greater than a number $\sim \hbar$.  This principle is logically
distinct from 
the uncertainty principle usually 
so-called~\cite{WodArtKel,Raymer,Hilgevoord,Uncertainty}.

In ref.~\cite{Appleby} we have shown that Eqs.~(\ref{eq:  RetErrRelB}--\ref{eq:  ErrDisRelB}) hold for many other
simultaneous measurement processes, apart from the Arthurs-Kelly process.

The distinction between the two different
aspects of a quantum mechanical measurement
process---the retrodictive or determinative
aspect \emph{versus} the predictive or
preparative one---has been discussed
by numerous authors, as has the unavoidable
perturbation of the system by the 
measurement~\cite{Bus4,Prug1,Bus2,Bus3,Mart2,
Margenau,Hilgevoord}.  The quantities
$\RErr x$, $\RErr p$, $\PErr x$, $\PErr p$, $\Dist x$, $\Dist p$,
and the inequalities relating them,
provide a
a convenient numerical characterisation of
these features.  However, it ought to be stressed that the
characterisation is not complete.  It is, for instance, clearly impossible
to give an exhaustive description of the change in the state
of the system by only specifying two numbers.  The 
values of the six quantities
defined in this section encapsulate some important properties
of the measurement process.  It is not to be supposed that 
they  encapsulate every significant property.

The method of describing statistical quantities
in terms of variances, and mean square 
values---the method adopted here, in other words---is subject to certain
limitations.  In recent years there has accordingly been some interest
in devising alternative approaches.  One approach is that
involving parameter-based uncertainty relationships~\cite{Hilgevoord,Braunstein2}.
Another approach is that involving entropic uncertainty
relationships~\cite{Bus3,Mart1,Mart2,Buzek2}.  It would be interesting
to see if either or both of these approaches could be used to 
develop the formalism of this paper.

It will be observed, that the retrodictive and predictive rms errors
enter Eqs.~(\ref{eq:  RetErrRelB}--\ref{eq:  ErrDisRelB}) in a completely
symmetric manner.  As we will see the two kinds of error can also be
fixed independently of one another (in the case of the Arthurs-Kelly
process).  With the appropriate choice of initial apparatus state it is
possible to arrange for the retrodictive errors to be small while
the predictive errors are large; or for the predictive errors to be 
small while the retrodictive errors are large; or for the retrodictive and predictive
errors both to be small.  In particular, there exists an initial apparatus
state for which $\RErr x = \PErr x$, $\RErr p = \PErr p$ and
for which the products $\RErr x \RErr p$ and $\PErr x \PErr p$ both take 
the minimum value of $\frac{\hbar}{2}$.  We will refer to such measurements
as completely optimal.  They are discussed in Section~\ref{sec:  complete}.

It should, however, be stressed that this symmetry between the 
retrodictive and predictive aspects of a simultaneous measurement
process disappears when one turns to the physical interpretation
of the quantities $\RErr x$, $\RErr p$, $\PErr x$, $\PErr p$,
as discussed in Section~5 of ref.~\cite{Appleby}.  One should also
bear in mind the fact just mentioned, that these quantities only afford
a partial characterisation of the measurement process.
When we refer to them as the rms errors of retrodiction and prediction
we mean, strictly and precisely, no more, and no less than what is  stated
in Section~5 of ref.~\cite{Appleby}.

The subject of retrodiction and prediction, and the relationship between
them, raises some deep conceptual questions which have been the subject
of discussion ever since the formulation of the modern theory of 
quantum mechanics in the mid-1920's~\cite{HeisBook}. It appears to us
that the methods developed in ref.~\cite{Appleby}, and in the present paper, 
might be used to gain some additional insight into these questions.
However, there is clearly a great deal more work which needs to be done.
\section{The Distribution of Measured Values}
\label{sec:  DistMeasVal}
We see from Eq.~(\ref{eq:  ErrDisExp}) that the error and disturbance operators only depend on
the apparatus observables $\MxOp$, $\PxOp$, $\MpOp$, $\PpOp$.  It follows that the rms errors and
disturbances are independent of the initial system state $\ket{\psi}$.  We also see that the
operators $\Exi$, $\Exf$ constitute a complete commuting set for the apparatus state space, with 
conjugate momenta $- \Epi$, $\Epf$.  It will be convenient to work in terms of 
simultaneous eigenkets of $\Exi$ and $\Exf$, which we denote
$\ket{\Exiv,\Exfv}_{\Exi,\Exf}$.  They are related to the simultaneous eigenkets of
$\MxOp$ and $\PpOp$, denoted $\ket{\Mx,\Pp}_{\MxOp,\PpOp}$, by
\begin{equation*}
  \bket{\Exiv,\Exfv}_{\Exi,\Exf} = \bket{\tfrac{1}{2}\left(\Exiv+\Exfv\right),\left(\Exiv-\Exfv\right)}_{\MxOp,\PpOp}
\label{eq:  ExiEpiRep}
\end{equation*}
We now use this equation, Eq.~(\ref{eq:  HatUDef}) and the Baker-Campbell-Hausdorff identity
to deduce
\begin{align*}
&   \bsubmatB{x,\Mx,\Pp}{\hat{U}}{\psi \otimes \pt}{\xOp,\MxOp,\PpOp}{3}
\\
& \hspace{0.25 in}
  = \bsubmatB{  x,\Mx,\Pp
            }{  \exp \left(-\tfrac{i}{2\hbar} \PpOp \PxOp\right)
                \exp\left(-\tfrac{i}{\hbar} \PpOp \pOp\right)
                \exp\left(-\tfrac{i}{\hbar} \PxOp \xOp\right)
            }{  \psi \otimes \pt
            }{  \xOp,\MxOp,\PpOp }{3}
\\
& \hspace{0.25 in}
  = \bsuboverlapB{x - \Pp}{\psi}{\xOp}{3} \ 
    \bsuboverlapB{\left(\Mx+\Pp - x\right), \left(\Mx - x\right)}{\pt}{\Exi,\Exf}{3}
\end{align*}
Taking Fourier transforms we get
\begin{align}
&   \bsubmatB{x,\Mx,\Mp}{\hat{U}}{\psi \otimes \pt}{\xOp,\MxOp,\MpOp}{3}
\notag
\\
& \hspace{0.25 in}
=   \sqrt{\frac{1}{h}}
    \int d x' \, \exp\left[\tfrac{i}{\hbar}  \Mp \left(x-x'\right) \right] \ 
                  \bsuboverlapB{\left(\Mx - x'\right), \left(\Mx - x\right)}{\pt}{\Exi,\Exf}{3} \ 
                  \bsuboverlapB{x'}{\psi}{\xOp}{3}
\label{eq:  FinTotWFunc}
\end{align}
We can now calculate $\rho \left( \Mx,\Mp \right)$, the probability density function describing the
result of the measurement:
\begin{align}
&   \rho \left( \Mx,\Mp \right)
\notag
\\
& \hspace{0.25 in}
= \int dx \left|  \bsubmatB{x,\Mx,\Mp}{\hat{U}}{\psi \otimes \pt}{\xOp,\MxOp,\MpOp}{3} \right|^2
\notag
\\
& \hspace{0.25 in}
= \frac{1}{h} \int dx' dx'' \, \exp \left[ \tfrac{i}{\hbar} \Mp \left( x'' - x'\right)\right]
                                  \mat{\Mx - x'}{ \hat{\rho}_{\Ei}  }{\Mx - x'' }
                                  \overlap{x'}{\psi} \overlap{\psi}{x''}
\label{eq:  RhoTermsPsi}
\end{align}
where $\hat{\rho}_{\Ei}$ is the reduced initial apparatus state density matrix corresponding to the
pair $\Exi$, $-\Epi$:
\begin{equation*}
    \bmat{  \Exiv^{\vphantom{'}}   }{  \hat{\rho}_{\Ei}  }{\Exiv' }
=   \int d \Exfv^{\vphantom{'}} \ 
        \bsuboverlapB{\Exiv^{\vphantom{'}},\Exfv^{\vphantom{'}}}{\pt}{\Exi, \Exf}{3} \;
        \boverlap{\pt}{\Exiv',\Exfv^{\vphantom{'}}}_{\Exi,\Exf}
\end{equation*}
Let $W_{\mathrm{sy,i}}$ be the Wigner function describing the initial state of the system, and let
$W_{\Ei}$ be the Wigner function corresponding to $\hat{\rho}_{\Ei}$:
\begin{equation*}
\begin{split}
        W_{\mathrm{sy,i}} (x,p) 
    & = \frac{1}{h} \int dy \, 
                       \exp\left( \tfrac{i}{\hbar} py\right) \, 
                       \overlap{x-\tfrac{1}{2} y}{\psi} \overlap{\psi}{x+ \tfrac{1}{2} y} \\
        W_{\Ei} \left( \Exiv, \Epiv\right)
    & = \frac{1}{h} \int dy \,
                       \exp\left( - \tfrac{i}{\hbar} \Epiv y \right) \, 
                       \mat{\Exiv - \tfrac{1}{2} y}{\hat{\rho}_{\Ei}}{\Exiv + \tfrac{1}{2} y}
\end{split}
\end{equation*}
Then the distribution of measured values can be written
\begin{equation}
    \rho \left( \Mx,\Mp \right)
=   \int dx dp \, W_{\Ei} \left( \Mx - x, \Mp - p \right) W_{\mathrm{sy, i}} (x,p)
\label{eq:  DistResTermsInit}
\end{equation}
With a suitable choice for the distribution of retrodictive errors, it is possible 
to obtain any member of the class of 
operational phase space distributions, discussed
by Davies~\cite{Davies}, W\'{o}dkiewicz~\cite{Operational} 
and others~\cite{Bus4,Ban,Lalovic,Halliwell}.

The fact, that the Arthurs-Kelly process can be used to obtain any 
member of this class of
distributions is shown in
Leonhardt~\cite{Leon3} (also see Ban~\cite{Ban}). 
The novelty of the derivation just given consists in the
fact that we have shown that the distribution of 
measured values depends only on the distribution of
retrodictive errors, and is independent of the distribution of 
predictive errors.

In certain cases the convolution in 
Eq.~(\ref{eq:  DistResTermsInit}) can 
be inverted~\cite{Bus3,Lalovic,Wuen}.  This
means, that the original state can be
reconstructed from the measured probability
distribution
\emph{provided} that the latter is known with
perfect accuracy---a fact which is
sometimes expressed by saying that
the measurement is informationally
complete~\cite{Bus4,Bus3}.  However, it should
be observed that the fact is of less practical
usefulness than may initially appear due to the
amplification of statistical errors which
occurs when one tries actually to carry out the
inversion using real experimental
data~\cite{Leon4}.

The right hand side of Eq.~(\ref{eq:  DistResTermsInit}) only depends on the distribution of 
retrodictive errors.  If, on the other hand, one wants to relate $\rho \left(\Mx,\Mp\right)$
to the final state of the system, then one needs to consider the distribution
of predictive errors.  We confine ourselves to the case when the
initial apparatus state factorises:
\begin{equation}
    \overlap{\Exiv,\Exfv}{\pt} = \overlap{\Exiv}{\phi_{\Ei}} \overlap{\Exfv}{\phi_{\Ef}}
\label{eq:  ApWveFncFactor}
\end{equation}

Suppose that the pointer positions $\Mx$, $\Mp$ are recorded to be in a region
$\mathscr{R}$.  Let $\hat{\rho}_{\mathrm{sy,f}}$ be the reduced density matrix representing the 
state of the system immediately after the measurement.  Then
\begin{equation*}
   \bmat{x}{\hat{\rho}_{\mathrm{sy,f}}}{x'}
=  \frac{1}{p_{\mathscr{R}}}
   \int_{\mathscr{R}} d \Mx d \Mp \,
       \bmat{x,\Mx,\Mp}{\hat{U}}{\psi \otimes \pt}  \bmat{\psi \otimes \pt}{\hat{U}}{x',\Mx,\Mp}
\end{equation*}
where $p_{\mathscr{R}}$ is the probability of finding $\Mx$ and $\Mp$ in the region
$\mathscr{R}$:
\begin{equation*}
    p_{\mathscr{R}}  =  \int_{\mathscr{R}} d \Mx d \Mp \, \rho \left( \Mx, \Mp \right)
\end{equation*}
Using Eqs.~(\ref{eq:  FinTotWFunc}), (\ref{eq:  RhoTermsPsi}) and~(\ref{eq:  ApWveFncFactor})
we obtain
\begin{align}
&  \bmat{x}{  \hat{\rho}_{\mathrm{sy,f}}  }{x'}
\notag
\\
& 
=  \frac{1}{ p_{\mathscr{R}} }
   \int_{\mathscr{R}} d \Mx d \Mp \,
      \exp\left[ \tfrac{i}{\hbar} \Mp \left( x - x' \right) \right] \,
      \rho \left( \Mx, \Mp \right) \ 
      \bsuboverlapB{\Mx - x}{\phi_{\Ef}}{\Exf}{3} \  \boverlap{\phi_{\Ef}}{\Mx - x'}_{\Exf} 
\label{eq:  FinSyRedDM}
\end{align}
Let $W_{\mathrm{sy,f}}$ be the Wigner function describing the final state of the system,
and let $W_{\Ef}$ be the Wigner function corresponding to the state $\ket{\phi_{\Ef}}$:
\begin{equation*}
\begin{split}
        W_{\mathrm{sy, f}} (x, p) 
    & = \frac{1}{h} \int dy \, \exp\left(\tfrac{i}{\hbar} p y\right) 
                          \mat{x-\tfrac{1}{2} y}{\hat{\rho}_{\mathrm{sy,f}}}{x+\tfrac{1}{2} y} \\
        W_{\Ef} \left( \Exfv, \Epfv \right)
    & = \frac{1}{h} \int dy \, \exp\left(\tfrac{i}{\hbar} \Epfv y\right)
                           \overlap{\Exfv -\tfrac{1}{2} y }{\phi_{\Ef}} \, \overlap{\phi_{\Ef}}{\Exfv+\tfrac{1}{2} y}
\end{split}
\end{equation*}
Then Eq.~(\ref{eq:  FinSyRedDM}) becomes
\begin{equation}
    W_{\mathrm{sy, f}} (x, p) 
=   \frac{1}{p_{\mathscr{R}} } 
    \int_{\mathscr{R}} d \Mx d \Mp \,
        W_{\Ef} \left( \Mx - x, \Mp - p \right) \rho \left( \Mx, \Mp \right)
\label{eq:  FinTermsDistRes}
\end{equation}
Eq.~(\ref{eq:  DistResTermsInit}) shows how the distribution of retrodictive errors can be used
to express $\rho$ in terms of $W_{\mathrm{sy, i}}$.
Eq.~(\ref{eq:  FinTermsDistRes}) shows how the distribution of predictive errors can be used 
to express $W_{\mathrm{sy, f}} $ in terms of $\rho $.

If $\mathscr{R}$ is a sufficiently small region surrounding the point 
$(\Mx, \Mp)$
\begin{equation*}
           W_{\mathrm{sy, f}} (x, p) 
\approx    W_{\Ef} \left( \Mx - x, \Mp - p \right) 
\end{equation*}
We see, that with a suitable choice for the distribution of predictive errors,
the Arthurs-Kelly process can be used to prepare the system in any desired pure state.
\section{Retrodictively Optimal Measurements}
\label{sec:  RetOpt}
Suppose that the measurement maximises  the degree of retrodictive accuracy:
\begin{equation}
    \RErr x \RErr p = \frac{\hbar}{2}
\label{eq:  OptRetErrRel}
\end{equation}
Define the quantity $\lambda_{\mathrm{i}}$ by
\begin{equation*}
    \RErr x = \frac{\lambda_{\mathrm{i}}}{\sqrt{2}} \hspace{0.5 in}
    \RErr p = \frac{\hbar}{\sqrt{2} \lambda_{\mathrm{i}}}
\end{equation*}
We will refer to $\lambda_{\mathrm{i}}$ as the retrodictive spatial resolution.

The necessary and sufficient condition for Eq.~(\ref{eq:  OptRetErrRel}) to be true is that
the initial apparatus wave function be of the form
\begin{equation}
  \overlap{\Exiv,\Exfv}{\pt} 
= \left(\frac{1}{\pi \lambda_{\mathrm{i}}^2} \right)^{\frac{1}{4}}
  \exp\left( - \frac{1}{2 \lambda_{\mathrm{i}}^2} \Exiv^2 \right) \phi_{\mathrm{f}} \left( \Exfv \right)
\label{eq:  RetOptAppWveFnc}
\end{equation}
where $\phi_{\mathrm{f}} \left( \Exfv \right)$ is an arbitrary normalised function.
$\phi_{\mathrm{f}}$  determines the errors of prediction.  The fact that it is arbitrary
means, that requiring the measurement to be retrodictively optimal places no
constraint on the predictive accuracy.  The two kinds of error are completely independent.

In the $\PxOp$, $\MpOp$-representation Eq.~(\ref{eq:  RetOptAppWveFnc}) takes the form
\begin{equation*}
  \overlap{\Px,\Mp}{\pt} 
= \left( \frac{\lambda_{\mathrm{i}}^2}{\pi \hbar^2} \right)^{\frac{1}{4}}
  \exp\left[ - \frac{\lambda_{\mathrm{i}}^2}{2 \hbar^2} \left( \Mp - \tfrac{1}{2} \Px \right)^2\right]
  \widetilde{\phi}_{\mathrm{f}} \left( \Mp + \tfrac{1}{2} \Px \right)
\end{equation*}
where $\widetilde{\phi}_{\mathrm{f}}$ is the Fourier transform of $\phi_{\mathrm{f}}$:
\begin{equation*}
    \widetilde{\phi}_{\mathrm{f}} \left( \Epfv \right)
=   \sqrt{\frac{1}{h}} \int d \Exfv \, \exp\left(- \frac{i}{\hbar} \Epfv \Exfv \right) \, 
\phi_{\mathrm{f}} \left( \Exfv \right)
\end{equation*}
We recognise the wave function considered by Stenholm~\cite{Stenholm}.

Using Eq.~(\ref{eq:  DistResTermsInit}) we find, for the probability distribution of measured values,
\begin{equation*}
    \rho \left( \Mx, \Mp \right)
=   Q_{\mathrm{i},\lambda_{\mathrm{i}}} \left( \Mx, \Mp \right)
\end{equation*}
where $Q_{\mathrm{i},\lambda_{\mathrm{i}}}$ is the initial system
state Husimi function~\cite{Husimi,Reviews}:
\begin{equation*}
    Q_{\mathrm{i},\lambda_{\mathrm{i}}} \left(\Mx, \Mp\right)
=   \frac{2}{h}
    \int dx dp \, \exp\left[ - \frac{1}{\lambda_{\mathrm{i}}^2} \left( \Mx - x\right)^2 
                      - \frac{\lambda_{\mathrm{i}}^2}{\hbar^2} \left( \Mp - p \right)^2
                    \right]
                  W_{\mathrm{sy,i}} (x,p)
\end{equation*}
The fact that the
Husimi function gives the distribution of
measured values for the case of maximal accuracy
was first shown by Ali and 
Prugove\v{c}ki~\cite{Prug2b}, working in terms of the approach based
on POVM's and unsharp observables.  The specialisation of this result to the
case of the Arthurs-Kelly process was first discussed by
Busch~\cite{Bus2}.

Finally, let us calculate the disturbances in
this case.  We have
\begin{equation}
\begin{split}
    \left(\Dist x\right)^2
& = \int d \Exiv d \Exfv\, \left(\Exiv-\Exfv\right)^2 \left|\overlap{\Exiv,\Exfv}{\pt}\right|^2
  = \frac{\lambda_{\mathrm{i}}^2}{2} + \left( \PErr x\right)^2
\\
    \left(\Dist p\right)^2
& = \int d \Epiv d \Epfv\, \left(\Epiv-\Epfv\right)^2 \left|\overlap{\Epiv,\Epfv}{\pt}\right|^2
  = \frac{\hbar^2}{2 \lambda_{\mathrm{i}}^2} + \left( \PErr p\right)^2
\label{eq:  RetOptDis}
\end{split}
\end{equation}
Using the predictive error relationship, Eq.~(\ref{eq:  PreErrRelB}), we deduce
\begin{equation*}
  \Dist x \, \Dist p \ge \hbar
\end{equation*}
\section{Predictively Optimal Measurements}
\label{sec:  PreOpt}
We next consider measurements which maximise the  predictive accuracy:
\begin{equation}
    \PErr x \, \PErr p = \frac{\hbar}{2}
\label{eq:  PreOptErrRel}
\end{equation}
Define the quantity $\lambda_{\mathrm{f}}$ by
\begin{equation*}
    \PErr x = \frac{\lambda_{\mathrm{f}}}{\sqrt{2}} \hspace{0.5 in} 
    \PErr p = \frac{\hbar}{\sqrt{2} \lambda_{\mathrm{f}}}
\end{equation*}
We will refer to $\lambda_{\mathrm{f}}$ as the predictive spatial resolution.
The necessary and sufficient condition for Eq.~(\ref{eq:  PreOptErrRel}) to be true is that 
$\overlap{\Exiv,\Exfv}{\pt}$ be of the form
\begin{equation}
    \overlap{\Exiv,\Exfv}{\pt}
=   \left( \frac{1}{\pi \lambda_{\mathrm{f}}^2} \right)^{\frac{1}{4}}
    \exp\left(- \frac{1}{ 2 \lambda_{\mathrm{f}}^2} \Exfv^2 \right) \phi_{\mathrm{i}} \left( \Exiv \right)
\label{eq:  PreOptAppWveFnc}
\end{equation}
Suppose that the final pointer positions are recorded to be in the region $\mathscr{R}$.
In view of Eq.~(\ref{eq:  FinSyRedDM}) the final system state reduced density matrix is given by
\begin{equation}
    \hat{\rho}_{\mathrm{sy,f}}
=   \frac{1}{p_{\mathscr{R}}} 
    \int_{\mathscr{R}} d \Mx d \Mp \, \rho \left( \Mx, \Mp \right) \;
    \bket{\coh{\Mx}{\Mp}{\lambda_{\mathrm{f}}}} \, \bbra{\coh{\Mx}{\Mp}{\lambda_{\mathrm{f}}}}
\label{eq:  PreOptFinSysSte}
\end{equation}
where $\bket{\coh{\Mx}{\Mp}{\lambda_{\mathrm{f}}}}$ is the coherent state with wave function
\begin{equation*}
   \boverlap{x}{\coh{\Mx}{\Mp}{\lambda_{\mathrm{f}}}}
=  \left( \frac{1}{\pi \lambda_{\mathrm{f}}^2} \right)^{\frac{1}{4}}
   \exp\left[ - \frac{1}{2 \lambda_{\mathrm{f}}^2} \left( x - \Mx \right)^2 
       + \frac{i}{\hbar} \Mp x - \frac{i}{2 \hbar} \Mp \Mx
      \right]
\end{equation*}
Let $P_{\mathrm{f},\lambda_{\mathrm{f}}}$ be the anti-Husimi function describing the final
state of the system (the $P$-representation of Glauber and Sudarshan).  We have~\cite{Reviews,Glauber}
\begin{equation}
    \hat{\rho}_{\mathrm{sy,f}}
=   \int d\Mx d\Mp \, P_{\mathrm{f},\lambda_{\mathrm{f}}} \left(\Mx, \Mp \right) \;
                      \bket{\coh{\Mx}{\Mp}{\lambda_{\mathrm{f}}}} \, \bbra{\coh{\Mx}{\Mp}{\lambda_{\mathrm{f}}}}
\label{eq:  AHFncDef}
\end{equation}
Comparing Eqs.~(\ref{eq:  PreOptFinSysSte}) and~(\ref{eq:  AHFncDef}) we deduce
\begin{equation*}
    P_{\mathrm{f},\lambda_{\mathrm{f}}} \left(\Mx, \Mp \right)
=   \begin{cases} 
         \frac{1}{p_{\mathscr{R}}} \rho \left(\Mx, \Mp \right) & \qquad \text{if $ \left(\Mx, \Mp \right) \in \mathscr{R}$}\\
         0 & \qquad \text{otherwise}
    \end{cases}
\end{equation*}
If $\mathscr{R}$ is a sufficiently small region surrounding the point $\left(\Mx,\Mp\right)$, then
the system is approximately in the state $\bket{\coh{\Mx}{\Mp}{\lambda_{\mathrm{f}}}}$ after the measurement:
\begin{equation*}
         \hat{\rho}_{\mathrm{sy,f}}
\approx  \bket{\coh{\Mx}{\Mp}{\lambda_{\mathrm{f}}}} \, \bbra{\coh{\Mx}{\Mp}{\lambda_{\mathrm{f}}}}
\end{equation*}
The reader may easily verify that
\begin{equation*}
  \Dist x \, \Dist p \ge \hbar
\end{equation*}
as in the case of a retrodictively optimal process.
\section{Completely Optimal Measurements}
\label{sec:  complete}
Suppose that the measurement is both retrodictively optimal at spatial resolution $\lambda_{\mathrm{i}}$,
and predictively optimal at spatial resolution $\lambda_{\mathrm{f}}$.  In view of 
Eqs.~(\ref{eq:  RetOptAppWveFnc}) and~(\ref{eq:  PreOptAppWveFnc}) the initial apparatus
wave function must be
\begin{equation*}
    \overlap{\Exiv,\Exfv}{\pt} 
=   \left(\pi \lambda_{\mathrm{i}} \lambda_{\mathrm{f}} \right)^{-\frac{1}{2}}
    \exp\left( - \frac{1}{2 \lambda_{\mathrm{i}}^2} \Exiv^2  - \frac{1}{2 \lambda_{\mathrm{f}}^2} \Exfv^2 \right)
\end{equation*}
We have from Eq.~(\ref{eq:  RetOptDis})
\begin{equation*}
    \Dist x \, \Dist p 
=   \frac{\hbar}{2} \left( 2 + \frac{\lambda_{\mathrm{f}}^2}{\lambda_{\mathrm{i}}^2} 
                             + \frac{\lambda_{\mathrm{i}}^2}{\lambda_{\mathrm{f}}^2}
                          \right)^{\frac{1}{2}}
\ge \hbar
\end{equation*}
The necessary and sufficient condition for this expression to achieve its lower bound
is that the retrodiction and prediction both be at the 
same spatial resolution:  $\lambda_{\mathrm{i}} = \lambda_{\mathrm{f}} = \lambda$, say.
We then have, in the $\MxOp$,$\MpOp$ representation
\begin{equation*}
    \overlap{\Mx,\Mp}{\pt}
=   \frac{2}{\sqrt{h}} \exp\left(- \frac{1}{\lambda^2} \Mx^2 - \frac{\lambda^2}{\hbar^2} \Mp^2\right)
\end{equation*}
which is the wave function considered by Arthurs and Kelly~\cite{ArthursKelly}.
With this choice of $\ket{\pt}$ the process produces the least amount of disturbance
consistent with maximal accuracy.  It might therefore be described as a completely
optimal process.

It is interesting to note, however, that $\lambda_{\mathrm{i}}$ and $\lambda_{\mathrm{f}}$
are completely independent.  One could,
for instance, have $\lambda_{\mathrm{i}} \rightarrow 0$ and $\lambda_{\mathrm{f}} \rightarrow \infty$---so that
the measurement gives an almost perfectly accurate retrodiction of position, whilst at the same time
preparing the system in an approximate eigenstate of momentum. 
Of course, the disturbances would then be
very large.
\section{Minimally Disturbing Measurements}
\label{sec:  MinDis}
It is possible to make $\Dist x \Dist p$ smaller than $\hbar$, provided that one is willing
to accept some loss of accuracy.  In this section we address the question:
what is the greatest accuracy which can be achieved for a given level of disturbance?
We confine ourselves to the case when the retrodictive and predictive errors are equal:
\begin{equation}
    \RErr x = \PErr x  \hspace{0.5 in} \text{and} \hspace{0.5 in} \RErr p = \PErr p
\label{eq:  EqErrs}
\end{equation}
We assume that the product of disturbances is given by
\begin{equation}
  \Dist x \, \Dist p = \hbar e^{-\eta}
\label{eq:  MinProdDist}
\end{equation}
for some $\eta \ge 0$.  We then ask:  what is the least  value of the product
$\RErr x \, \RErr p = \PErr x \, \PErr p$ subject to this constraint?  
And:  what is the probability distribution of
measured values when the lower bound is achieved?

It is convenient to define
\begin{equation}
    \Ex = \tfrac{1}{2} \left( \Exi + \Exf \right) \hspace{0.5 in} \text{and} \hspace{0.5 in}
    \Ep = \tfrac{1}{2} \left( \Epi + \Epf \right)
\label{eq:  AveErrDef}
\end{equation}
We also have
\begin{equation}
    \Dx =  \Exi - \Exf  \hspace{0.5 in} \text{and} \hspace{0.5 in}
    \Dp =  \Epi - \Epf 
\label{eq:  ErrToDisEqs}
\end{equation}
Consequently
\begin{equation*}
    \bcomm{\Ex}{\Dp} = \bcomm{\Dx}{\Ep} = - i \hbar
\end{equation*}
all other commutators between  $\Ex$, $\Dx$, $\Ep$, $\Dp$ being zero.
We see that $\Ex$, $\Dx$ constitute a complete commuting set of apparatus observables, with
conjugate momenta $-\Dp$, $-\Ep$.  In particular
\begin{equation}
\begin{split}
    \bmat{\pt}{\Ex^2}{\pt} \, \bmat{\pt}{\Dp^2}{\pt} & \ge \frac{\hbar^2}{4} \\
    \bmat{\pt}{\Dx^2}{\pt} \, \bmat{\pt}{\Ep^2}{\pt} & \ge \frac{\hbar^2}{4}
\end{split}
\label{eq:  AveErrDistRels}
\end{equation}
It follows from Eqs.~(\ref{eq:  AveErrDef}) and~(\ref{eq:  ErrToDisEqs})
\begin{equation*}
\begin{split}
        \left( \RErr x\right)^2  
    & = \bmat{\pt}{\Ex^2}{\pt} + \frac{1}{4} \bmat{\pt}{\Dx^2}{\pt} + \bmat{\pt}{\Ex \Dx}{\pt}\\
        \left( \PErr x\right)^2  
    & = \bmat{\pt}{\Ex^2}{\pt} + \frac{1}{4} \bmat{\pt}{\Dx^2}{\pt} - \bmat{\pt}{\Ex \Dx}{\pt}
\end{split}
\end{equation*}
In view of Eq.~(\ref{eq:  EqErrs}) we must have
\begin{equation*}
   \bmat{\pt}{\Ex \Dx}{\pt} = 0
\end{equation*}
and
\begin{equation}
    \left( \RErr x\right)^2 = \left( \PErr x\right)^2
=   \bmat{\pt}{\Ex^2}{\pt} + \frac{1}{4} \bmat{\pt}{\Dx^2}{\pt}
\label{eq:  RerrTermsAveErr}
\end{equation}
Similarly
\begin{equation}
    \left( \RErr p\right)^2 = \left( \PErr p\right)^2
=   \bmat{\pt}{\Ep^2}{\pt} + \frac{1}{4} \bmat{\pt}{\Dp^2}{\pt}
\label{eq:  PerrTermsAveErr}
\end{equation}
In view of Eq.~(\ref{eq:  MinProdDist}) we can choose $\lambda$ such that
\begin{equation*}
\begin{split}
    \Dist x & = \lambda \, \exp\left(-\tfrac{\eta}{2}\right) \\ 
    \Dist p & = \frac{\hbar}{\lambda} \, \exp\left(-\tfrac{\eta}{2}\right)
\end{split}
\end{equation*}
In view of Eq.~(\ref{eq:  AveErrDistRels}) we must then have
\begin{equation}
\begin{split}
    \bmat{\pt}{\Ex^2}{\pt} & \ge \frac{\lambda^2}{4} \, e^{\eta} \\
    \bmat{\pt}{\Ep^2}{\pt} & \ge \frac{\hbar^2}{4 \lambda^2} \, e^{\eta}
\end{split}
\label{eq:  AveErrBounds}
\end{equation}
Inserting these results in Eqs.~(\ref{eq:  RerrTermsAveErr}) and~(\ref{eq:  PerrTermsAveErr})
gives
\begin{equation*}
\begin{split}
        \left( \RErr x\right)^2 = \left( \PErr x\right)^2
  & \ge \frac{\lambda^2}{2} \cosh \eta \\
        \left( \RErr p\right)^2 = \left( \PErr p\right)^2
  & \ge \frac{\hbar^2}{2 \lambda^2} \cosh \eta 
\end{split}
\end{equation*}
whence
\begin{equation*}
     \RErr x \, \RErr p = \PErr x \, \PErr p \ge \frac{\hbar}{2} \cosh \eta
\end{equation*}
which is the desired inequality.  

The product of errors achieves its lower bound if and only if the 
lower bounds set by  Eq.~(\ref{eq:  AveErrBounds}) are achieved, so that 
$\ket{\pt}$ is a minimum uncertainty state with respect to the pairs 
$\Ex$, $- \Dp$ and $\Dx$, $-\Ep$.  In the $\Ex$, $\Dx$-representation
\begin{equation*}
    \overlap{\epsilon_{\mathrm{X}},\delta_{\mathrm{X}}}{\pt}
=   \frac{1}{\sqrt{\pi} \lambda}
    \exp\left[ - \frac{1}{\lambda^2} 
                 \left( e^{- \eta} \epsilon_{\mathrm{X}}^2 + \tfrac{1}{4 } e^{\eta} \delta_{\mathrm{X}}^2 \right)
      \right]
\end{equation*}
Transforming to the $\Exi$,$\Exf$-representation we find
\begin{equation*}
    \overlap{\Exiv,\Exfv}{\pt} 
=   \frac{1}{\sqrt{\pi} \lambda} 
    \exp\left[ - \frac{1}{2 \lambda^2} \left(\cosh \eta \, \Exiv^2 
               - 2 \sinh \eta \, \Exiv \Exfv 
               + \cosh \eta \, \Exfv^2 \right)
        \right]
\end{equation*}
Using Eq.~(\ref{eq:  DistResTermsInit}) we obtain the probability distribution of measured values
\begin{align*}
&   \rho \left( \Mx, \Mp \right)
\\
& \hspace{0.25 in}
=   \frac{2}{h \cosh \eta}
    \int dx dp \,  
           \exp\left[ - \frac{1}{\cosh \eta} \left( \frac{1}{\lambda^2} \left( \Mx - x\right)^2 
                                           + \frac{\lambda^2}{\hbar^2} \left( \Mp - p \right)^2
                                      \right)
                \right]
          W_{\mathrm{i,sy}} (x,p)
\end{align*}
This is a smeared Wigner function, of the kind proposed by Cartwright~\cite{Lalovic,Halliwell,Cartwright}.  For
$\eta = 0$ it reduces to the Husimi function.  As $\eta$ increases 
$W_{\mathrm{sy,i}}$ is smoothed over increasingly large regions of phase space---in agreement
with the fact, that the larger $\eta$, 
the less accurate the measurement.
\section{Conclusion}
In this paper we have only considered the 
Arthurs-Kelly process.  It would be
interesting to know whether the results
obtained generalise to some of the other
measurement processes which have been
discussed in the 
literature~\cite{Leon1,LeonUncert,Torma,Leon2,Leon3}.  We
hope to return to this question in a future
publication.
\section*{Appendix:  the Definition of the Retrodictive Errors}
\label{sec:  appendix}
In  Eq.~(\ref{eq:  RetErrDef}) we defined the rms errors of retrodiction 
$\RErr x$, $\RErr p$ in terms of the retrodictive error operators
$\Exi$, $\Epi$.  
The purpose of this appendix is to indicate the relationship
between these quantities and the definition of the
measurement inaccuracies which was given by 
Ali and Prugove\v{c}ki~\cite{Bus4,Prug2b}.

Consider a measurement process in which a system, 
with position $\xOp$ and momentum $\pOp$
interacts with an apparatus, characterised by two pointer observables 
$\MxOp$, $\MpOp$.  Let $\psi$ and $\pt$ be the initial states of the system
and apparatus respectively.  Let $\hat{U}$ be the unitary 
evolution operator describing the interaction, so that the
final state of the system$+$apparatus is
$\hat{U} \ket{\psi \otimes \pt}$ (in the Schr\"{o}dinger picture).
We no longer confine ourselves to the  
case of the Arthurs-Kelly process, so $\hat{U}$ 
is not assumed to have the particular form 
specified by Eq.~(\ref{eq:  HatUDef}).

For a given choice of initial apparatus state $\pt$ we have
\begin{equation}
  \bmat{x, \Mx, \Mp}{\hat{U}}{\psi \otimes \pt}
= \int dx' \,
      K(x,\Mx,\Mp;x') \, \boverlap{x'}{\psi}
\label{eq:  UTermsK}
\end{equation}
for some kernel $K$ (the choice of $K$ being dependent on the 
choice of $\pt$).  The unitarity of $\hat{U}$ implies
\begin{equation}
  \int dx d\Mx d\Mp \,
     K(x,\Mx,\Mp; x_1) \, K^{*} (x,\Mx,\Mp;x_2)
=  \delta (x_1-x_2)
\label{eq:  unitarity}
\end{equation}
The distribution of measured values is given by
\begin{align*}
  \rho (\Mx, \Mp)
& = \int dx \, \left| \bmat{x, \Mx, \Mp}{\hat{U}}{\psi \otimes \pt}\right|^2
\\
& =  \int dx dx' \vphantom{x}_1 dx'\vphantom{x}_2 \,
        K(x,\Mx,\Mp; x' \vphantom{x}_1) \, K^{*} (x,\Mx,\Mp; x' \vphantom{x}_2) \,
        \boverlap{x' \vphantom{x}_1}{\psi}
        \boverlap{\psi}{x' \vphantom{x}_2}
\end{align*}
or, in terms of the $p$-representation wavefunction,
\begin{equation*}
  \rho (\Mx, \Mp)
=  \int dp d p' \vphantom{p}_1 d p' \vphantom{p}_2 \,
       \tilde{K}(p, \Mx, \Mp; p' \vphantom{p}_1)\,
       \tilde{K}^{*}(p, \Mx, \Mp; p' \vphantom{p}_2)\,
        \boverlap{p' \vphantom{p}_1}{\psi}
        \boverlap{\psi}{p' \vphantom{p}_2}
\end{equation*}
where
\begin{equation*}
    \tilde{K}(p, \Mx, \Mp; p')
=   \frac{1}{h^2}
    \int dx dx' \,
       \exp\left[ \frac{i}{\hbar} (p' x' - p x) \right]
       K(x,\Mx,\Mp; x')
\end{equation*}

The marginal distributions can be written
\begin{equation*}
\begin{split}
  \int d\Mp \, \rho(\Mx,\Mp)
& = \int d x' \vphantom{x}_1 \, d x' \vphantom{x}_2 \,
     f_{\mathrm{X}} (\Mx; x' \vphantom{x}_1, x' \vphantom{x}_2) \,
        \boverlap{x' \vphantom{x}_1}{\psi}
        \boverlap{\psi}{x' \vphantom{x}_2}
\\
  \int d\Mx \, \rho(\Mx,\Mp)
& = \int d p' \vphantom{p}_1 \, d p' \vphantom{p}_2 \,
     f_{\mathrm{P}} (\Mp; p' \vphantom{p}_1, p' \vphantom{p}_2) \,
        \boverlap{p' \vphantom{p}_1}{\psi}
        \boverlap{\psi}{p' \vphantom{p}_2}
\end{split}
\end{equation*}
where
\begin{equation}
\begin{split}
   f_{\mathrm{X}} (\Mx; x' \vphantom{x}_1, x' \vphantom{x}_2)
& =  \int d x d \Mp \,
      K(x,\Mx,\Mp; x' \vphantom{x}_1 ) \,
      K^{*} (x,\Mx,\Mp; x' \vphantom{x}_2 )
\\
   f_{\mathrm{P}} (\Mp; p'\vphantom{p}_1, p' \vphantom{p}_2)
& =  \int d p d \Mx \,
      \tilde{K} (p,\Mx,\Mp; p' \vphantom{p}_1 ) \,
      \tilde{K}^{*} (p,\Mx,\Mp; p' \vphantom{p}_2 )
\end{split}
\label{eq:  fDefs}
\end{equation}
Ali and Prugove\v{c}ki~\cite{Prug2b} confine themselves to  the class of processes for which
$f_{\mathrm{X}}$ and $f_{\mathrm{P}}$ have the 
form
\begin{equation*}
\begin{split}
   f_{\mathrm{X}} (\Mx; x' \vphantom{x}_1, x' \vphantom{x}_2)
& =  \chi_{\mathrm{X}} (\Mx,  x' \vphantom{x}_1) \,  \delta(x' \vphantom{x}_2 - x' \vphantom{x}_1)
\\
   f_{\mathrm{P}} (\Mp; p' \vphantom{p}_1, p' \vphantom{p}_2)
& =  \chi_{\mathrm{P}} (\Mp,  p' \vphantom{p}_1) \,  \delta(p' \vphantom{x}_2 - p' \vphantom{x}_1)
\end{split}
\end{equation*}
for some pair of functions
$\chi_{\mathrm{X}}$, $\chi_{\mathrm{P}}$, 
and they focus on the even more restricted class for which
\begin{equation}
\begin{split}
   f_{\mathrm{X}} (\Mx; x' \vphantom{x}_1,  x' \vphantom{x}_2)
& =  \chi_{\mathrm{X} 0} (\Mx -  x' \vphantom{x}_1) \,  \delta(x' \vphantom{x}_2 - x' \vphantom{x}_1)
\\
   f_{\mathrm{P}} (\Mp; p' \vphantom{p}_1 , p' \vphantom{p}_2)
& =  \chi_{\mathrm{P} 0} (\Mp -  p' \vphantom{p}_1) \,  \delta(p' \vphantom{x}_2 - p' \vphantom{x}_1)
\end{split}
\label{eq:  PrugConv}
\end{equation}
for some pair of functions
$\chi_{\mathrm{X}0}$, $\chi_{\mathrm{P}0}$,
This assumption is valid 
in the case of the Arthurs-Kelly process.  However, there does not seem to be
any reason to
expect it to be true generally.

If $f_{\mathrm{X}}$ and $f_{\mathrm{P}}$ do satisfy the 
condition of Eq.~(\ref{eq:  PrugConv}), 
the marginal probability distributions for $\Mx$ and $\Mp$ can be
written as convolutions of the initial system state probability distributions for 
$x$ and $p$:
\begin{equation*}
\begin{split}
  \int d\Mp \, \rho(\Mx,\Mp)
& = \int d x'\,
     \chi_{\mathrm{X}0} (\Mx- x') \,
        \left|\boverlap{x'}{\psi}\right|^2
\\
  \int d\Mx \, \rho(\Mx,\Mp)
& = \int d p'\,
     \chi_{\mathrm{P}0} (\Mp - p') \,
        \left|\boverlap{p'}{\psi}\right|^2
\end{split}
\end{equation*}
Ali and Prugove\v{c}ki~\cite{Bus4,Prug2b} define the measurement inaccuracies in terms 
of the widths of these convolutions:
\begin{equation}
\begin{split}
   \sigma_{\mathrm{X}} 
& = \left( \int dx \, x^2 \, \chi_{\mathrm{X}0} (x)
         - \left(\int dx \, x \, \chi_{\mathrm{X}0}(x) \right)^2\right)^{\frac{1}{2}}
\\
   \sigma_{\mathrm{P}} 
& = \left( \int dp \, p^2 \, \chi_{\mathrm{P}0} (p)
         - \left(\int dp \, p \, \chi_{\mathrm{P}0}(p) \right)^2\right)^{\frac{1}{2}}
\end{split}
\label{eq:  PrugErr}
\end{equation}

Let us now compare these definitions with the definitions of
the rms errors of retrodiction
used in this paper.  Referring back to Eq.~(\ref{eq:  ErrDisOps})
we see
\begin{align*}
& \bmat{\psi \otimes \pt}{\Exi^2}{\psi \otimes \pt}
\\
& \hspace{0.5 in}
 = \bmat{\psi \otimes \pt
      }{  \hat{U}^{\dagger} \MxOp^2 \hat{U} 
         + \hat{U}^{\dagger} \MxOp \hat{U} \xOp
         +\xOp \hat{U}^{\dagger} \MxOp \hat{U}
         + \xOp^2
     }{ \psi \otimes \pt}
\end{align*}
Using Eqs~(\ref{eq:  UTermsK}) and~(\ref{eq:  fDefs}) we find
\begin{align*}
    \bmat{\psi \otimes \pt
      }{  \hat{U}^{\dagger} \Mx^2 \hat{U} 
     }{ \psi \otimes \pt}
& = \int d \Mx d \Mp dx \,
        \Mx^2 
        \left| \bmat{x,\Mx,\Mp}{\hat{U}}{\psi \otimes \pt} \right|^2
\\
& = \int d \Mx d{x'}_1 d{x'}_2 \,
        \Mx^2 f_{\mathrm{X}} (\Mx; {x'}_1, {x'}_2)
        \overlap{{x'}_1}{\psi} \overlap{\psi}{{x'}_2}
\end{align*}
and
\begin{align*}
&   \bmat{\psi \otimes \pt
      }{  \hat{U}^{\dagger} \Mx \hat{U} \xOp
     }{ \psi \otimes \pt}
\\
& \hspace{0.3 in}
= \int d \Mx d \Mp dx \,
      \Mx \, \bmat{x,\Mx,\Mp}{\hat{U} \xOp}{\psi \otimes \pt} \,
      \bmat{\psi \otimes \pt}{\hat{U}^{\dagger}}{x,\Mx,\Mp}
\\
& \hspace{0.3 in}
= \int d \Mx d{x'}_1 d{x'}_2 \,
      \Mx \, {x'}_1 \,
      f_{\mathrm{X}} (\Mx; {x'}_1, {x'}_2)
      \overlap{{x'}_1}{\psi} \overlap{\psi}{{x'}_2}
\end{align*}
Similarly
\begin{align*}
&  \bmat{\psi \otimes \pt
      }{  \xOp \hat{U}^{\dagger} \Mx \hat{U}
     }{ \psi \otimes \pt}
\\
& \hspace{0.3 in}
= \int d \Mx d{x'}_1 d{x'}_2 \,
      \Mx \, {x'}_2 \,
      f_{\mathrm{X}} (\Mx; {x'}_1, {x'}_2)
      \overlap{{x'}_1}{\psi} \overlap{\psi}{{x'}_2}
\end{align*}
Lastly
\begin{align*}
  \bmat{\psi \otimes \pt
      }{  \xOp^2
     }{ \psi \otimes \pt}
& =  \int dx' \, {x'}\vphantom{x}^2 \left| \boverlap{x'}{\psi} \right|^2
\\
& = \int d \Mx d{x'}_1 d{x'}_2 \,
      {x'}_1 \, {x'}_2 \,
      f_{\mathrm{X}} (\Mx; {x'}_1, {x'}_2)
      \overlap{{x'}_1}{\psi} \overlap{\psi}{{x'}_2}
\end{align*}
where we have used the fact
\begin{equation*}
  \int d\Mx \, f_{\mathrm{X}} (\Mx; {x'}_1, {x'}_2)
= \delta ({x'}_1 - {x'}_2)
\end{equation*}
as follows from the unitarity condition, Eq.~(\ref{eq:  unitarity}).
Putting these results together gives
\begin{equation*}
   \left(\RErr x\right)^2
=  \int d \Mx d{x'}_1 d{x'}_2 \,
       (\Mx - {x'}_1) (\Mx - {x'}_2) \, f_{\mathrm{X}} (\Mx; {x'}_1,{x'}_2) \,
       \boverlap{{x'}_1}{\psi} \boverlap{\psi}{{x'}_2}
\end{equation*}
In the same way we can derive
\begin{equation*}
   \left(\RErr p\right)^2
=  \int d \Mp d{p'}_1 d{p'}_2 \,
       (\Mp - {p'}_1) (\Mp - {p'}_2) \, f_{\mathrm{P}} (\Mp; {p'}_1,{p'}_2) \,
       \boverlap{{p'}_1}{\psi} \boverlap{\psi}{{p'}_2}
\end{equation*}
These equations hold quite generally.
If the functions $f_{\mathrm{X}}$, $f_{\mathrm{P}}$ are of the 
form given by Eq.~(\ref{eq:  PrugConv}), then
\begin{equation*}
\begin{split}
      \left(\RErr x\right)^2
& =  \int d x \, x^2 \, \chi_{\mathrm{X}0} (x)
\\
      \left(\RErr p\right)^2
& =  \int d p \, p^2 \, \chi_{\mathrm{P}0} (p)
\end{split}
\end{equation*}
Comparing these equations with Eq.~(\ref{eq:  PrugErr})
we see, that the only difference is, that Ali and Prugove\v{c}ki work 
in terms of the standard deviations, whereas we use the rms values.
Our definition essentially coincides with theirs, therefore, for the
class of measurement processes which they consider (and which includes
the Arthurs-Kelly process discussed in this paper).

Our definition of the retrodictive errors applies to a  larger class
of measurement processes than the definition of Ali and Prugove\v{c}ki.
It also has other advantages, since it brings out the connection with
the retrodictive error operators.  In ref.~\cite{Appleby} we  
used this connection to answer the question of principle raised by Uffink, and to prove
the error-error and error-disturbance relationships of 
Eqs.~(\ref{eq:  RetErrRelB}-\ref{eq:  ErrDisRelB}).  In the present paper we
saw how the expansion in terms of eigenstates of the retrodictive
and predictive error operators provides a very convenient way of 
analysing the mathematical properties of the Arthurs-Kelly process.

\end{document}